\documentclass[twocolumn,aps,prd,amsmath,amssymb,preprintnumbers,longbibliography]{revtex4-1}
\usepackage{amsmath} \usepackage{amsfonts} \usepackage{amssymb}
\usepackage{bbm}
\usepackage{epsfig}
\usepackage{graphics}
\usepackage{graphicx}
\usepackage{titlesec}
\usepackage{mathtools}
\usepackage{environ}
\usepackage{dsfont}
\usepackage{version}
\usepackage[colorlinks=true]{hyperref}
\hypersetup{urlcolor=black,linkcolor=black,citecolor=black}
\usepackage{natbib}
\usepackage[toc,page]{appendix}

\usepackage{dsfont}

\makeatletter
\renewcommand*\env@matrix[1][c]{\hskip -\arraycolsep
  \let\@ifnextchar\new@ifnextchar
  \array{*\c@MaxMatrixCols #1}}
\makeatother

\makeatletter										
\def\p@subsection{\thesection .\,} 
\makeatother                                       

\usepackage{textalpha} 

\newcommand{\be}{\begin{equation}}
\newcommand{\ee}{\end{equation}}
\newcommand{\ba}{\begin{align}}
\newcommand{\ea}{\end{align}}
\newcommand{\nn}{\nonumber}

\newcommand{\gl}{\big(}
\newcommand{\gr}{\big)}

\newcommand{\bel}[1]{\be\label{#1}}

\titleformat{\subsection}[block]{\normalfont\bfseries}{\thesubsection.}{1ex}{}
\titlespacing{\subsection}{0pt}{10pt}{1pt}[0pt]
\titleformat*{\section}{\large\bfseries}
\renewcommand{\thesubsection}{\arabic{subsection}}

\pdfoutput=1 

\usepackage{todonotes}
\usepackage{braket}

\newcommand{\2}{^{2}}
\newcommand{\vi}{V_{\infty}}
\newcommand{\vm}{V_{-\infty}}

\newcommand{\qq}[1]{``#1"}
\newcommand{\vp}{{\varphi}}
\newcommand{\eps}{\varepsilon}
\newcommand{\wtil}{\tilde w}
\newcommand{\util}{\tilde u}
\newcommand{\rhotil}{\tilde\rho}
\newcommand{\ubar}{\bar u}
\newcommand{\kapbar}{\bar\kappa}

\definecolor{refkey}{rgb}{0,0,1}
\definecolor{labelkey}{rgb}{0,1,0}

\begin{document}


\title{\LARGE The quantum gravity connection between inflation \\and quintessence.}

\author{C. Wetterich}

\affiliation{Institut  f\"ur Theoretische Physik\\
Universit\"at Heidelberg\\
Philosophenweg 16, D-69120 Heidelberg}

\begin{abstract}
Inflation and quintessence can both be described by a single scalar field. The cosmic time evolution of this cosmon field realizes a crossover from the region of an ultraviolet fixed point in the infinite past to an infrared fixed point in the infinite future. This amounts to a transition from early inflation to late dynamical dark energy, with intermediate radiation and matter domination. The scaling solution of the renormalization flow in quantum gravity connects the two fixed points. It provides for the essential characteristics of the scalar potential needed for the crossover cosmology and solves the cosmological constant problem dynamically. The quantum scale symmetry at the infrared fixed point protects the tiny mass of the cosmon and suppresses the cosmon coupling to atoms without the need of a non-linear screening mechanism, thereby explaining apparent issues of fine tuning. For a given  content of particles the scaling solution of quantum gravity is a predictive framework for the properties of inflation and dynamical dark energy.

\end{abstract}

\maketitle
\bigskip
\noindent


Inflation~\cite{STA,GUT,MUK,ALI,SWI} and dynamical dark energy or quintessence~\cite{Wetterich_1988,RP1,RP2} can both be mediated by the cosmic evolution of a scalar field that is neutral with respect to the gauge symmetries of the standard model of particle physics. It is a natural idea to identify these fields~\cite{SPO,PEVI,PERO,DIVA,GIO, BRMA}. In short, the inflaton becomes the cosmon of dynamical dark energy. In this note we investigate the question if such scenarios of ``quintessential inflation" or ``cosmon inflation" can be understood in terms of some fundamental physics properties. We will focus on predictions and constraints from quantum gravity. This will shed light on some issues of fine tuning or naturalness that at first sight seem to be present within such scenarios. We focus on general aspects and refer to refs~\cite{CWCI,CWIQM,RUCW,HMSS,HMSS2,WHMSS,HASO,BR} for concrete models and their compatibility with observation. We present our questions and answers in ten points.

\bigskip\noindent\textbf{(1) Overall picture}
\smallskip

\begin{figure}[h]
\centering\includegraphics[width=0.48\textwidth]{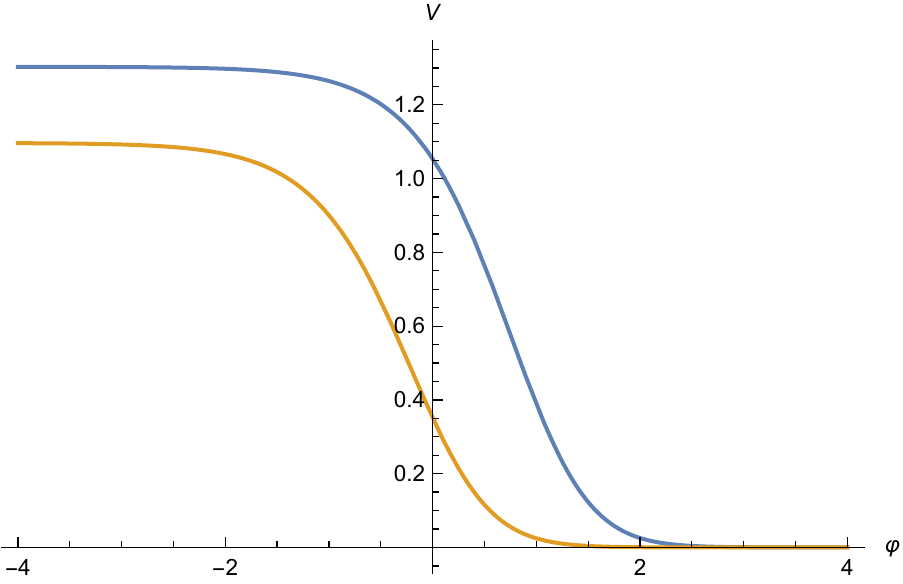}
\caption{General shape of the scalar potential $V(\varphi)$ for quintessential inflation. Both the scalar field $\varphi$ and the potential $V$ are in Planck units. The two curves correspond to models discussed later.}\label{fig:1}
\end{figure}

Both inflation and dynamical dark energy require a scalar potential $V(\varphi)$ that is almost flat for values of the scalar field $\varphi$ in the two regions which are relevant during inflationary cosmology and the present dark energy dominated cosmology. The rough form required for $V(\varphi)$ is depicted in Fig.~\ref{fig:1}.
The almost constant value $V_{-}$ for large negative $\varphi$ determines the Hubble parameter during inflation as $H^{2}=V_{-}/(3M^{2})$, with $M=2.44\cdot 10^{18} GeV$ the Planck mass. 
On the other hand, some value $V_{+}$ in the tail for large positive $\varphi$ can be identified with the present dark energy density of the Universe, $V_{+}\approx (2\cdot 10^{-3} eV)^{4}$. The two flat tails are connected by a crossover region with a more rapid variation of $V(\varphi)$. After inflation the Universe has to be heated, producing the particles present in the radiation dominated epoch and the associated entropy. The heating depends on the coupling of $\varphi$ to the fields for elementary particles in the standard model and possibly beyond.
We will not address this issue in the present note and refer to refs~\cite{RUCW,BR}. We focus on the question if there is a natural origin of the two flat tails connected by a crossover region.

\medskip\noindent\textbf{(2) Single scalar field}
\smallskip

The two regions for large negative and positive $\varphi$ look rather different: the values of $V_{+}$ and $V_{-}$ differ by many orders of magnitude. One possible explanation for this qualitative difference could be that the inflaton and cosmon are actually two different fields. The cosmic evolution may follow a valley (relative minimum) in the common potential for the two fields and one may parametrize the location within the valley by $\varphi$. This could formally connect a region of negative $\varphi$, which is dominated by the inflaton field $\varphi_{\textup{inf}}$, to the region for large $\varphi$ dominated by the cosmon field $\varphi_{\textup{cos}}\,$.

This two-field setting is not what we discuss in this note. We are rather interested in ``single field models" where $\varphi$ has a common origin and meaning for the whole range shown in Fig.~\ref{fig:1}. Two questions become obvious: The first asks why a single field should be responsible for the dark energy density both for the inflationary and the present epoch. Our answer is an identification of $\varphi$ with a fundamental scalar field related to quantum scale symmetry. The second question asks for an explanation of the two apparently rather different tails of the potential.

\medskip\noindent\textbf{(3) Naturalness and fine tuning}
\smallskip

Let us denote $V_{\infty}=V(\varphi\to \infty) $ and$V_{-\infty}=V(\varphi\to -\infty) $. For $V_{\infty}\neq 0$ the model contains a tiny non-zero dimensionless parameter $V_{\infty }/ V_{-\infty}$. Without an explanation such a tiny parameter may be considered as unnatural. The obvious way to avoid a small non-zero parameter is given by $V_{\infty}=0$. 

There may be several contributions to $\vi$. They have to be ``fine tuned" such that their sum vanishes for $\vi =0$, or results in a tiny value for non-zero $\vi / \vm$. We want to find a natural mechanism that explains $\vi =0$, and therefore forces individual contributions to sum up to zero. Such a mechanism solves the so called ``cosmological constant problem'' \cite{WEICC}. We propose that this mechanism is rooted in the fixed point behavior of the renormalization flow. Typical flow generators or $\beta$-functions can have many different contributions from the fluctuations of different fields. Nevertheless, at a fixed point the flow generators vanish, such that all contributions add up precisely to zero. If the dynamics of the scalar field drives it for $\varphi\rightarrow \infty$ towards a range of fields for which a fixed point is realized, this fixed point will govern the behavior of the potential. 
In our case we need that the fixed point value for $\vi$ vanishes. Fixed points occur in the flow of couplings or functions with a renormalization scale $k$. The proposed mechanism requires that the flow with $k$ can be mapped to a change of the scalar field value $\varphi$. We we will see below how this is realized. 

Furthermore, for dynamical dark energy the present mass of the cosmon is very small, typically of the order of the present Hubble parameter $H\approx 10^{-33} eV$. In the presence of quantum fluctuations such a tiny mass may again appear as a fine-tuning problem. Quantum scale symmetry~\cite{CWQS} at a fixed point solves this issue. At the infrared fixed point, which is reached for $\varphi \rightarrow\infty$, quantum scale symmetry is an exact global symmetry of the effective action, while being broken spontaneously by a non-zero value of a scalar field $\chi$, which is related to $\varphi$ by a simple field transformation (see below).  On the one side this spontaneous symmetry breaking is responsible for the observed non-zero particle masses which are proportional to $\chi$. At the same time, the spontaneous breaking of the global scale symmetry induces a Goldstone boson - the dilaton - which is precisely massless at the fixed point. For large finite $\varphi$ in the vicinity of the fixed point quantum scale symmetry is only approximate, leading to an almost massless pseudo-Goldstone boson. The cosmon is the pseudo-Goldstone boson of the spontaneously broken approximate scale symmetry, which explains its tiny mass. The shape of the potential and the cosmon mass are directly related, since the latter involves the second derivative $\partial^{2}V/\partial\varphi^{2}$. 

Finally, for dynamical dark energy the scalar field $\varphi$ changes its value even in the present epoch. Since we do not invoke any non-linear screening mechanism, this requires that the coupling of $\varphi$ to the atoms of ordinary matter must be sufficiently small. Otherwise one could observe a time-variation of fundamental ``constants" or an apparent violation of the equivalence principle~\cite{CWCNC,DAPO,CWCMAV,CHIB,UZ,DZ,CWPQT,DPV,CWTFC}. In the presence of quantum fluctuations the smallness of the cosmon-atom coupling may again appear as a problem of fine tuning. We will see that exact quantum scale symmetry results in vanishing cosmon-atom couplings. This gives a natural explanation for small couplings~\cite{CWQS} if the present value of $\varphi$ is in the vicinity of a fixed point with the associated quantum scale symmetry. 

\medskip\noindent\textbf{(4) Variable gravity}
\smallskip

In a quantum field theory for the metric and a scalar field the scalar potential is not the only relevant function of the scalar field. Also the coefficient of the curvature scalar $R$ in the quantum effective action, which is related to an effective Planck mass, will depend on the value of a scalar field. One expects non-zero non-minimal couplings $\sim \xi \chi^{2} R$, where $\chi$ may be the Higgs field or, in our case, a singlet field related to $\varphi$. We will therefore discuss the effective action for variable gravity\cite{CWVG},
\begin{equation}\label{1}
\Gamma=\int_{\chi}\sqrt{g}\Big{\lbrace} -\frac{1}{2}F(\chi)R+\frac{1}{2} K(\chi)\partial^{\mu}\chi\partial_{\mu}\chi+U(\chi)\Big{\rbrace}\, .
\end{equation}
With three $\chi$-dependent functions $F$, $K$ and $U$ this is the most general form for a derivative expansion in second order in the derivatives for the coupled system of the metric field $g_{\mu\nu}$ and a scalar field $\chi$. We require here diffeomorphism symmetry. In our notation $g=\det (g_{\mu\nu})$ provides a factor $i$ in $\sqrt{g}$.

Variable gravity is a rather modest version of modified gravity for which the effective squared Planck mass $F$ depends on $\chi$~\cite{Wetterich_1988}. It belongs to the general class of scalar-tensor theories. The Brans-Dicke theory~\cite{RDMPI} would be obtained for $F=\chi^{2}$, $K=K_{0}$ and $U=0$ if the particle masses are constant. In contrast, for our setting it is important that also the masses $m_{p}$ in the particle physics sector (not explicitly specified in eq.~\eqref{1}) depend on $\chi$~\cite{CWCNC,Wetterich_1988}. We will find for large $\chi$ a behavior $F\sim \chi^{2}$, $m_{p}\sim \chi$, such that the ratio particle-mass / effective Planck mass $m_{p}/\sqrt{F}$ is independent of $\chi$. In this region all particle mass ratios as well as the dimensionless couplings of the standard model are independent of $\chi$. 
This behavior will be dictated by quantum scale symmetry.  From the point of view of phenomenology the $\chi$-dependence of the particle masses constitutes a crucial difference to Brans-Dicke theory or extensions with a cosmological constant~\cite{BER,FOR,NWE} which do not allow a realistic matter dominated epoch for constant $m_{p}$. Only the behavior $m_{p}\sim\chi$ allows the cosmology of variable gravity to be compatible with observation.

By a Weyl scaling~\cite{HWGE} $g_{\mu\nu}=(M^{2}/F)g'_{\mu\nu}$, together with a rescaling of the scalar field $\varphi=4M\ln(\chi/k)$, the effective action takes the form
\begin{equation}
\label{2}
\Gamma=\int_{\chi}\sqrt{g}\bigg{\lbrace} -\frac{M^{2}}{2}R'+\frac{1}{2} Z(\varphi)\partial^{\mu}\varphi\partial_{\mu}\varphi+V(\varphi)\bigg{\rbrace}\, .
\end{equation}
The constant Planck mass $M$ has been introduced here by the variable transformation rather than being a fundamental parameter. Covariant derivatives, the curvature scalar $R'$ and $\sqrt{g'}$ are now formed from the metric $g'_{\mu\nu}$ in the Einstein frame. The relation between the two frames is given by 
\begin{equation}
\label{3}
V(\varphi)=\frac{UM^{4}}{F^{2}}\, ,\quad Z(\varphi)=\frac{1}{16}\bigg{\lbrace}\frac{\chi^{2}K}{F}
+\frac{3}{2}\Big{(}\frac{\partial\ln F}{\partial\ln \chi}\Big{)}^{2}\bigg{\rbrace}\ .
\end{equation}

The exact field equations derived by variation of the quantum effective action are strictly equivalent~\cite{CWCNC,FCDDGE,DTTG,CEJFC,DCECG,CCCO,CWEU,PEEJF,JIQG,JFICIM,KFCMI} for eqs. \eqref{1} and \eqref{2} - the two frames are related by a simple variable transformation in differential equations. This equivalence is called ``field relativity"\cite{CWUWE}. We observe that the same transformations have to be performed in the particle physics sector, accompanied by suitable field transformations for the fields for fermions or other scalars in order to maintain a canonical form of their kinetic terms. Typical particle masses scale $m_{p}\rightarrow m_{p}'=\sqrt{M/F}\,m_{p}$ such that the ratios remain unaffected, $m_{p}^{\prime 2}/M^{2}=m_{p}^{2}/F^{2}$.

The phenomenology of cosmology is most easily discussed in the Einstein frame for which the Planck mass and particle masses are constant. On the other hand, the ``quantum frame"~\eqref{1} of variable gravity is more appropriate for understanding the role of quantum fluctuations, the renormalization flow of coupling, the fixed points and the associated quantum scale symmetry. The non-linear field transformation to the Einstein frame obscures many simple properties. This makes it rather difficult to understand the naturalness of dynamical dark energy and inflation in the Einstein frame. The discussion of this note will be centered on the quantum frame and the quantum effective action \eqref{1} for variable gravity.

\medskip\noindent\textbf{(5) Crossover cosmology}
\smallskip

\begin{figure}[h]
\centering\includegraphics[width=0.48\textwidth]{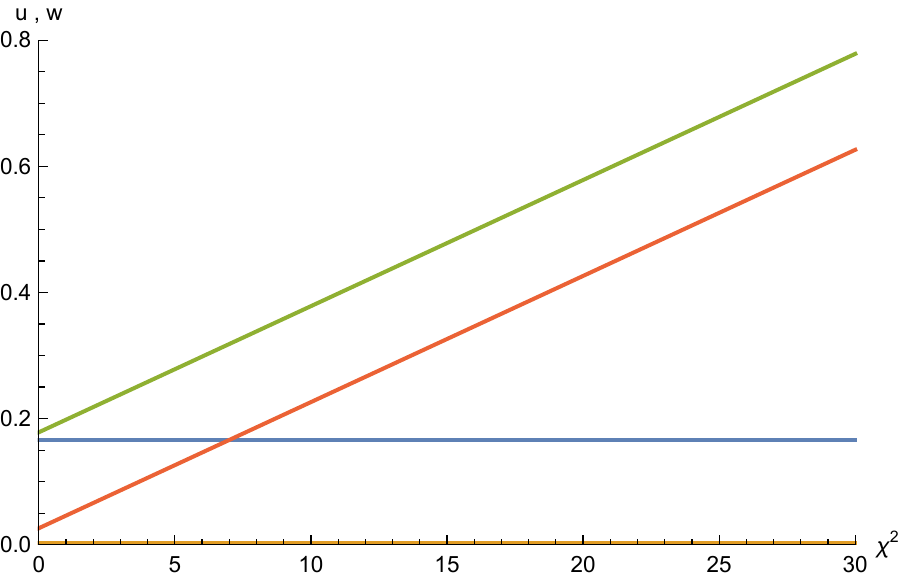}
\caption{Crossover functions. We plot $u= U/k^{4}$ (horizontal lines) and $w=F/(2k^{2})$ as functions of $\chi\2/M\2$. Parameters $u_{0}$ and $w_{0}$ are given by eqs.~\eqref{SS15} and~\eqref{SS17} with $\xi = 0.01$. The upper curves use $N_{V}=45$, $N_{F}=48$, $N_{S}=326$, as appropriate for SO(10) grand unification, and the lower curves are for $N_{V}=N_{F}=N_{S}=0$. }\label{fig:2}
\end{figure}

We first assume a simple form of the effective action which will be motivated by quantum gravity subsequently. For our rough ansatz the coefficients of the curvature scalar $F$ can be approximated by
\bel{CC4} 
F=2w_{0}k^2+\xi\chi^2\ ,
\ee
as depicted in Fig.~\ref{fig:2}.
For $\chi\to0$ the $\chi$-dependence becomes negligible, while for large $\chi$ the non-minimal scalar-gravity coupling $\xi$ becomes dominant. A crossover between the two limits occurs for $\chi^2/k^2=2w_{0}/\xi$. We consider a constant potential
\bel{CC5}
U=u_{0}k^4\ .
\ee
In eqs.~\eqref{CC4},~\eqref{CC5} we have factored out the (arbitrary) renormalization scale $k$ such that the constants $w_{0}$, $u_{0}$, $\xi$ as well as $K(\chi)$ are all dimensionless.

The potential in the Einstein frame takes for this simple crossover scenario an exponential form for large $\varphi$
\begin{align}
\label{CC6}
V=&\frac{u_{0}M^4}{\gl2w_{0}+\xi\exp\left( \frac{\vp}{2M} \right) \gr ^2}\nn\\
=&\frac{u_{0}M^4}{\xi^2}\Big[1+\frac{2w_{0}}{\xi}\exp\left(-\frac{\vp}{2M}\right)\Big]^{-2}\exp\left(-\frac\vp M\right)\ .
\end{align}
It vanishes for $\vp\to\infty$, $V_\infty=0$. The potential shown in Fig.~\ref{fig:1} is actually the Weyl-scaled form of the simple crossover situation~\eqref{CC4},~\eqref{CC5} shown in Fig.~\ref{fig:2}. (The upper curve is for $N_{V}=45$, $N_{F}=48$, $N_{S}=326$ as for a typical grand unified model with SO(10)-symmetry, the lower one for $N_{V}=N_{F}=N_{S}=0$.) The crossover between the flat tails for $\vp\to\pm\infty$ corresponds to the crossover in $F$. It seems much simpler to explain the crossover~\eqref{CC4},~\eqref{CC5} than the particular form in the Einstein frame. In particular, we note that in the quantum frame the cosmological constant $u_{0}k^4$ does not vanish. There is no need of exact cancellation of vacuum fluctuations - for $u_{0}$ of the order one $U$ has its natural value if $k$ is the only mass scale of the model. 

We observe that the scale $k$ is no longer present in the Einstein frame~\cite{HPW}. The value of this renormalization scale is arbitrary. The present dark energy density $V(\vp)$ is tiny for a large value of $\vp/M$ without any small dimensionless parameter of the model. This simple observation has been the basis of the first proposal of dynamical dark energy or quintessence~\cite{Wetterich_1988}. (For subsequent, more observation oriented work see refs.~\cite{RP1,RP2,CWCMAV,FTSW,FEJO,VL,CLW,CDS,LA1,LACQ}.)

For the region $\vp/M\gg2\ln(2w_{0}/\xi)$ the potential takes a simple exponential form $V/M^4=\exp\gl-(\vp+c)/M\gr$. If the epoch of inflation relevant for the observable fluctuation spectrum occurs in this region of a \qq{standard exponential potential}, the slow roll parameters $\eps$ and $\eta$ can be directly extracted from the \qq{kinetial} or \qq{wave function renormalization} $Z(\vp)$, according to~\cite{CWCI,CWIQM}
\bel{CC7}
\eps=\frac1{2Z(\vp)}\ ,\quad \eta=\frac1{Z(\vp)}\left(1+\frac M2\frac{\partial\ln Z(\vp)}{\partial\vp}\right)\ .
\ee
Assuming that for this region of $\vp$ the coupling $\xi$ varies only slowly one finds the approximate form
\bel{CC8}
Z(\vp)\approx\frac1{16}\left(\frac K\xi+6\right)\ .
\ee
Small values of the slow roll parameters require large values of $K/\xi$~\cite{CWCI}, see also ref.~\cite{GKLR,KDR}. For this scenario an end of inflation requires a $\vp$-dependence of $K/\xi$, which may be realized by a suitable form of $K(\chi)$. Inflation ends once $Z$ drops below one~\cite{CWCI,CWIQM}.

We will see below that $\chi\to0$ corresponds to an ultraviolet (UV)-fixed point. Quantum scale symmetry at this fixed point is realized for $K(\chi\to0)=\kappa (k/\chi)^\sigma$~\cite{CWPC}, with $\sigma$ the scalar anomalous dimension. For the infrared (IR)-fixed point for $\chi\to\infty$ quantum scale symmetry requires constant $K_\infty$. A particularly interesting possible IR-fixed point is the conformal fixed point with $K/\xi=-6$. In order to be consistent with both fixed points we make a first simple crossover ansatz,
\bel{CC9}
\frac K{16\xi}=\kappa\Big{(}\frac{ k}{\chi}\Big{)}^{\sigma}+\frac1{\alpha^2}-\frac38\ ,
\ee
resulting in
\bel{CC10}
Z(\vp)=\kappa\exp\left(-\frac{\sigma\vp}{4M}\right)+\frac1{\alpha^2(\vp)}\ .
\ee
For large $\vp$ the function $\alpha^{-2}(\vp)$ measures the distance form the conformal fixed point and may be slowly varying, reaching zero for $\vp\to\infty$. For large $\alpha$ the crossover in $K$ occurs for $(\chi/k)^{\sigma}\approx\kappa$.

Due to the possibility of a common multiplicative rescaling of $\chi$ and $k$ only the combinations $K/\xi$, $u_{0}/\xi^2$, $w_{0}/\xi$ matter. In the approximation of constant $\alpha$ our ansatz involves therefore the dimensionless parameters $\kappa$, $\sigma$, $\alpha$, $\wtil=2w_{0}/\xi$ and $v=u_{0}/w_{0}$. A shift in $\vp$ multiplies $\kappa$, $\wtil$ and $v$ by a common factor, that we may use to set $\wtil=1$ in eq.~\eqref{CC6} and to replace $u_{0}/\xi\2\to\ubar$, $\kappa\to\kapbar$. With this normalization of $\vp$ our ansatz has four free dimensionless parameters $\kapbar=\kappa\xi/(2w_{0})$, $\sigma$, $\alpha$ and $\ubar=u_{0}/(4w_{0}^2)$.

In the limit of the exponentially decaying potential the inflationary epoch lasts as long as $Z>1$ or $\exp(-\sigma\vp/4M)>1/\kapbar$. If we assume simultaneously $\exp(-\sigma\vp/4M)\ll1$ in order to have $\vp$ in the exponential tail of the potential~\eqref{CC6}, we have to require $\kapbar\gg1$. In this case the crossover triggering the end of inflation is the one in $K(\chi)$. We observe that at the end of inflation one has
\bel{CC11}
\frac {V_{f}}{M^4}=\ubar\kapbar^{-\tfrac{4}{\sigma}}=(2w_{0})^{\tfrac{4}{\sigma}-2}(\kappa\xi)^{-\tfrac{4}{\sigma}}u_{0}\ .
\ee
This can be a very small quantity for large enough $\kapbar$, reflecting in eq.~\eqref{CC6} the exponential suppression factor for $\vp\gg M$. A large value of $\kapbar$ can explain, and is needed for, the observed small amplitude of the primordial fluctuations~\cite{CWIQM,RUCW}. An alternative scenario would use the crossover in $F(\chi)$, as reflected in the corresponding crossover of the potential $V(\varphi)$, in order to trigger the end of inflation. Compatibility with a small amplitude of the primordial fluctuations would require in this case a very small value $u_{0}/w_{0}\2$. We will argue below that quantum gravity does not seem to allow a tiny $u_{0}/w_{0}\2$, and focus therefore on the crossover in $K$ for ending inflation.

The properties of the spectrum of primordial density fluctuations are directly related to the slow roll parameters. The spectral index $n$ of scalar fluctuations and the tensor to scalar ratio $r$ are therefore determined by $Z(\vp)$, with $\vp$ given by the value of the scalar field at a time corresponding to $N$ $e$-foldings before the end of inflation, at which the observable fluctuations are frozen,
\begin{align}
\label{PI1}
1-n&=6\eps-2\eta=\frac1{Z(\vp)}\left(1-M\frac{\partial\ln Z(\vp)}{\partial\vp}\right)\ ,\nn\\
r&=16\eps=\frac8{Z(\vp)}\ .
\end{align}
The number of $e$-foldings before the end of inflations obeys, with $\vp_f$ the value of $\vp$ at the end of inflation,
\bel{PI2}
N(\vp)=\frac1M\int_\vp^{\vp_f}\text{d}\vp'Z(\vp')\ .
\ee
Neglecting the term $\alpha^{-2}$ in eq.~\eqref{CC10} (see below) one finds the relation between $N$ and $\vp$ or $Z(\vp)$,
\bel{PI3}
N(\vp)=\frac4\sigma\gl Z(\vp)-1\gr\ ,
\ee
implying
\bel{PI4}
1-n=\frac{4+\sigma}{\sigma N+4}\ \quad r=\frac{32}{\sigma N+4}\ .
\ee
The bound $r<0.036$~\cite{BIKE} implies $\sigma N>885$, while $1-n\approx 0.035$~\cite{PLA} requires $\sigma$ near four. For a typical range $50<N<70$ one observes a clash between the two requirements and concludes that the simple ansatz~\eqref{CC10} is not compatible with observation.

It is possible to device other crossover-shapes of the kinetial $K(\chi)$ or $Z(\vp)$ that are compatible with present observation. Examples for very similar models can be found in refs.~\cite{CWIQM, RUCW, BR}. A discussion of the various possibilities for realistic inflation is not the point of this paper. We only indicate here that small $r$ requires a large value of $Z(\vp_{pf})$, where $\vp_{pf}$ indicates the value of $\vp$ relevant for the horizon crossing or freezing of the observable fluctuations,
\bel{PI5}
Z^{-1}(\vp_{pf})=\frac r8\lesssim\frac1{240}\ .
\ee
On the other hand, the relation
\bel{PI6}
M\frac\partial{\partial\vp}Z^{-1}(\vp_{pf})=\frac{\partial\ln Z}{\partial N}=1-n-\frac r8\approx\frac1{30}\ ,
\ee
indicates that $Z$ decreases at least by a factor two for a change in $\vp$ of the rough order $\Delta\vp/M\approx 1/8$. A realistic fluctuation spectrum seems to require a rather rapid crossover for $Z(\vp)$. We could formally account for this by including in eq.~\eqref{CC10} a suitable dependence of $\kappa$ and $\sigma$ on $\vp$. For a qualitative discussion we may continue with the approximation of constant $\kappa$ and $\sigma$.

For the large values of $\vp/M$ relevant for the post-inflationary cosmology one has $Z\approx\alpha^{-2}$. The field $\tilde{\sigma}=\vp/\alpha$ has a standard normalization of its kinetic term, with approximate potential
\bel{CC12}
V=\util M^4\exp\left(-\frac{\alpha\tilde{\sigma}}M\right)\ .
\ee
This is a standard potential for many models of quintessence~\cite{Wetterich_1988}. Cosmological scaling solutions obtain for large enough $\alpha$. In this case the term$\sim\alpha^{-2}$ in $Z$ plays only a small role during the inflationary epoch. Essentially, the parameters $\kappa$, $\sigma$ and $\util=u_{0}/\xi\2$ determine the dynamics of inflation, while $\alpha$ governs the behavior of dynamical dark energy. We will discuss the properties of dynamical dark energy in the points (8)-(10).

\medskip\noindent\textbf{(6) Scaling solution in (dilaton) quantum gravity}
\smallskip

We come now to a central point of this note, namely that scaling solutions in quantum gravity predict the qualitative form~\eqref{CC4},~\eqref{CC5} for the $\chi$-dependence of $F$ and $U$. In contrast, the crossover behavior of $K$ in eq.~\eqref{CC9} is not yet established. More precisely, $\xi$ can be a slowly varying function of $\chi$ with a constant positive value $\xi_\infty$ for $\chi\to\infty$, while a slowly varying $u=U/k^{4}$ is found to take constant values $u_0$ and $u_\infty$ for $\chi\to0$ and $\chi\to\infty$, interpolating smoothly between them.

If quantum gravity is a renormalizable quantum field theory, the functional flow equations~\cite{Wetterich_1993,RWEAEE,MR} have to admit a scaling solution~\cite{HPRW,HPW,CWQS,CWESPA,EIPAU,LPSW} for which the dimensionless combinations $U/k^4$, $F/k^2$ and $K$ become fixed functions of the dimensionless ratio $\chi^2/k^2$. This scaling solution permits to follow the flow to arbitrarily large $k$, corresponding to arbitrarily short length scales. At the ultraviolet fixed point for $k\to\infty$ nothing changes anymore, permitting to extrapolate the model to arbitrarily high momenta or short distances and to render thereby quantum gravity complete. This is a typical scenario of asymptotic safety~\cite{WEIN,MR,SOUM,DPER,RSAU,LAUR,NAPE,DELP,bonanno2020critical}, while asymptotic freedom~\cite{STE,FRATSE,AB} may also be possible~\cite{SWY} in the presence of higher order curvature terms. (We omit a discussion of the higher order curvature terms in this note because they play only a negligible role for the crossover cosmologies in the range relevant for observations.)

For the scaling solutions the dimensionless functions depend only on the dimensionless ratio $\rhotil=\chi^2/k^2$. This is the basic reason why the renormalization flow with $k$ is mapped directly to the $\chi$-dependence of the relevant couplings in the effective action, which translates in turn to cosmology by the dynamics of the evolution of $\chi$. The ultraviolet limit $k\to\infty$ at fixed $\chi$ can also be realized at fixed $k$ by $\chi\to0$, while the infrared limit $k\to0$ at fixed $\chi$ corresponds to $\chi\to\infty$ at fixed $k$. The crossover behavior of the functions $F$ and $K$ interpolates between the UV-fixed point properties for $\chi\to0$ and the IR-fixed point properties for $\chi\to\infty$.

For general renormalizable theories, including asymptotically safe of free quantum gravity, the renormalization flow departs from the ultraviolet fixed point and associated scaling solution as $k$ is lowered. This happens due to the presence of \qq{relevant parameters} of the flow. These relevant parameters turn into the free couplings of a model. A more radical perspective assumes \qq{fundamental scale invariance}~\cite{CWFSI}, for which the scaling solution holds for all $k$. This corresponds to some type of finite theory. A theory with fundamental scale invariance is very predictive since there are no more the free couplings associated to the relevant parameters. From a qualitative point of view the more general renormalization flow with relevant parameters is very similar to fundamental scale invariance if the scale $k_{f}$ of departure from a scaling solution is much smaller than the masses of all massive particles. This is realized for $k_{f}\approx10^{-3}\text{eV}$, except perhaps for neutrinos. We will focus here on the more predictive scheme of fundamental scale invariance.

For fundamental scale invariance the form of the functions $u(\rhotil)=U/k^4$, $w(\rhotil)=F/(2k^2)$ and $K(\rhotil)$, with $\rhotil=\chi^2/k^2$, are entirely determined by the scaling solution. The \qq{parameters} $u_{0}$, $w_{0}$, $\xi$, $\kappa$ and $\alpha$ of our crossover ansatz become predictable. Scaling solutions are very restricted since they have to solve a complex system of non-linear differential equations for the whole range of $\rhotil$. Finding them, establishing the crossover behavior and determining the effective parameters is a highly non-trivial computation. If successful, this will relate inflation and quintessence by the properties of fluctuations in quantum gravity.

The first steps in this direction are rather encouraging~\cite{HPRW,HPW,CWQS,CWESPA}. All candidate scaling solutions show the qualitative behavior~\eqref{CC4},~\eqref{CC5}. In particular, the scaling solution for the constants $u_0=u(\rhotil=0)$ and $u_\infty=u(\rhotil\to\infty)$ is understood rather easily. The flow equations for these couplings correspond to the ones for a type of cosmological constant, which may differ for the two limits. For the gauge invariant setting of the flow equations~\cite{CWPM} one finds for both limits~\cite{PRWY,CWESPA}
\begin{align}
\label{SS13} 
k\partial_kU=&\frac{k^4}{24\pi^2}\left(\frac5{1-v}+\frac1{1-v/4}+\frac34(N_{u}-4)\right)\ ,\nn\\
N_{u}=&N_S+2N_V-2N_F\ .
\end{align}
Here $N_S$, $N_V$ and $N_F$ are the numbers of effectively massless scalars, vector bosons and Weyl fermions, respectively. The parts independent of $N_{u}$ are the contribution of the metric fluctuations, with $v=u/w$. One has $v=v_0=u_0/w_0$ for the limit $\rhotil\to0$ and $v=v_\infty=0$ for the limit $\rhotil\to\infty$.

The scaling solution corresponds to fixed points of the flow equation
\bel{SS14}
k\partial_ku=k\partial_k\left(\frac U{k^4}\right)=-4u+4c_U\ ,
\ee
for which the r.h.s. vanishes. The fixed points are given by
\bel{SS15}
u_*=c_U=\frac1{192\pi^2}\left(\frac5{1-v}+\frac1{1-v/4}+\frac34(N_{u}-4)\right)\ .
\ee
The limit $\rhotil\to\infty$ corresponds to a renormalization scale $k$ much smaller than the effective Planck mass $\chi$. In this limit one deals with an effective low energy limit, as given by the standard model of particles plus an (almost) massless scalar field. For $\rhotil\to\infty$ only the photon ($N_V=1$) and the cosmon ($N_S=1$) contribute besides the metric, resulting in
\bel{SS16}
u_\infty=\frac7{256\pi^2}\ .
\ee
For $\chi\to0$ quantum scale symmetry is not spontaneously broken and all particles are massless. A computation of $u_0$ needs knowledge about the particles that play a role for momenta above the Planck mass. These may be the particles of the standard model, some grand unified model or even further extensions. One expects that $u_0$ differs from $u_\infty$, while being of a similar order of magnitude. The flow of $u(\rhotil)$ for intermediate $\rhotil$ is more complex, but it is not surprising that computations in models with a simplified particle content find a smooth interpolation between the two limits~\cite{HPRW,CWESPA}.

The flow equations for $w(\rhotil)$ are more involved, even though the gauge invariant flow equation yields a comparatively simple structure~\cite{CWMY} due to the decoupling between physical and gauge modes~\cite{CWMF}. The limit $\rhotil\to\infty$ is comparatively simple, since this covers an effective infrared theory for the metric, photons and the scalar $\chi$. One obtains indeed a scaling solution with $F\sim\xi_\infty\chi^2$~\cite{HPRW,HPW,CWQS,CWMY}. This is not surprising since $\xi$ corresponds to the dominant renormalizable coupling for $\chi\to\infty$. In contrast to $u_\infty$ the value of $\xi_\infty$ cannot be extracted from the asymptotic behavior alone - in this limit scaling solutions exist for arbitrary values of $\xi_\infty$. Restrictions on $\xi_\infty$ from the existence of scaling solutions arise since not all values allow for a continuation of the solution from the asymptotic region $\rhotil\to\infty$ to $\rhotil\to0$. This issue needs an understanding of the flow beyond the asymptotic region.

For $\rhotil\to0$ one can neglect $\xi$ in the leading behavior for $w(\rhotil)$. A computation similar to the one for $u_0$ yields a fixed point~\cite{CWMY} at
\ba\label{SS17}
&w_0=\frac{25}{128\pi^2(1-v_0)}+\frac1{192\pi^2}\left( N_{w}+\frac{43}6\right)\ ,\nn\\
&N_{w}=4N_{V}-N_{F}-N_{S}\ .
\end{align}
Taken together with the asymptotic behavior for large $\rhotil$ this establishes the qualitative crossover character of eq.~\eqref{CC4}. This qualitative behavior has indeed been found for all candidate scaling solutions in dilaton quantum gravity~\cite{HPRW,HPW}. We consider the qualitative behavior~\eqref{CC4},~\eqref{CC5} as a rather robust property of quantum gravity.

The characteristic size of $u_0$ and $w_0$ according to the scaling solution has an important consequence for inflationary cosmology. For the potential $V(\vp)$ shown in fig.~\ref{fig:1} an epoch of \qq{early inflation} occurs for values of $\vp$ in the region of the flat tail of the potential for $\vp\to-\infty$. During this early inflation the tensor to scalar ratio $r$ is predicted to be very small. The amplitude of the tensor fluctuations is given by the potential $V(\vp)/M^4$, which approaches for $\vp\to\infty$ the value $u_0/(4w_0^2)$. Stable gravity requires $w_0>0$, and inflation needs $u_0>0$, such that $v_0=u_0/w_0$ is positive. Despite the uncertainty for $N_u$, which refers to the particle content in the ultraviolet, the size of $u_0$ cannot be orders of magnitude smaller than $10^{-3}$. On the other hand, values of $w_0$ exceeding one do not seem to be plausible. (For the models shown in Fig.~\ref{fig:1} one has $u_0/(4w_0^2)$ of the order one.) Any value $V(\vp)/M^4>10^{-4}$ at the time of horizon crossing of the tensor fluctuations is much too large to be compatible with the observed bounds on their amplitude. The scaling solution requires that the primordial fluctuations should be frozen at a later time when $\vp$ has already reached the exponential tail of $V(\vp)$ for large $\vp\to\infty$. As discussed before, this needs large $Z(\vp)$ and a crossover in the kinetial in order to end inflation.

Much less is known about the scaling solution for $K(\rhotil)$. The behavior $K\sim\rhotil^{-\sigma/2}$ for $\rhotil\to0$, which is suggested by quantum scale symmetry, has not yet been investigated by explicit solutions of flow equations. For $\rhotil\to\infty$ one needs a setup which makes the fixed point at $K_\infty=-6\xi_\infty$ manifest. This fixed point is expected due to the enhanced conformal symmetry. 

\medskip\noindent\textbf{(7) Quantum scale symmetry}
\smallskip

In the UV-limit $k\to\infty$ and IR-limit $k\to0$ one expects fixed points for a theory with fundamental scale invariance~\cite{CWFSI}. At these fixed points quantum scale symmetry becomes exact~\cite{CWQS}. The existence of a UV-fixed point is required for a renormalizable theory of quantum gravity. Our crossover ansatz for $U$, $F$ and $K$ has to reflect this fixed point structure. At a fixed point the quantum effective action does not exhibit any mass scale. This includes the renormalization scale $k$. Thus $\Gamma$ has to become independent of $k$ for $\chi\to0$ and $\chi\to\infty$. The resulting global symmetry is quantum scale symmetry.

The absence of any mass scale has to hold for a suitable choice of fields. This choice may differ for the UV- and IR-fixed points. For the IR-fixed point we use for the effective action~\eqref{1} the metric field $g_{\mu\nu}$  and the scalar $\chi$. The potential $U=u_0k^4$ vanishes for $k\to0$, and $F=\xi\chi^2$ becomes independent of $k$ if $\xi$ reaches a constant value $\xi_\infty$. Also the kinetic term does not involve $k$ if $K(\chi\to\infty)=K_\infty$ reaches a constant. Our ansatz shows directly the invariance of $\Gamma$ under the global scale transformation $\chi\to\alpha\chi$, $g_{\mu\nu}\to\alpha^{-2}g_{\mu\nu}$.

Quantum scale symmetry would be compatible with a potential $U=\lambda\chi^4$ with constant $\lambda$. Such a term would lead to a cosmological constant $\lambda M^4$ in the Einstein frame~\cite{Wetterich_1988}. We find, however, that this term is not compatible with the scaling solution for quantum gravity. The absence of the term $\lambda\chi^4$ constitutes an important example how the requirement of a scaling solution of the system of non-linear differential flow equations restricts the possible couplings of a model. The scaling solution predicts a fixed point value $\lambda=0$. This prediction coincides with the general quantum gravity bound~\cite{CWIRG} for the increase of $U(\chi)$ for $\chi\to\infty$. For $F\sim\chi^2$ the potential is allowed to increase at most $\sim\chi^2$. Combined with the requirement of quantum scale symmetry this allows only $U(k\to0)=0$, as realized for $U=u_{\infty}k^4$.

The bound for the maximal increase of $U$ would still allow for a leading behavior $U\sim k^2\chi^2$ - a model that has often been studied in the past~\cite{CWVG,CWIQM,RUCW}. Scaling solutions of quantum gravity do not seem to exist for an asymptotic behavior $U\sim k^2\chi^2$. We have therefore not included this term in our ansatz. For late cosmology the behavior $U=u_\infty k^4$ or $U=\tilde\mu_\infty k^2\chi^2$ both lead to a very similar phenomenology~\cite{CWIQM}. In both cases the potential in the Einstein frame~\eqref{3} vanishes for $\chi\to\infty$, either $\sim\chi^{-4}$ or $\sim\chi^{-2}$.

The overall conclusion is rather striking. Quantum gravity solves the cosmological constant problem dynamically if cosmology is of a \qq{runaway type} where $\chi$ increases to infinity in the infinite future. This is the case for our setting. Fundamental scale invariance provides for an even stronger statement. The scaling solution of quantum gravity requires that the potential $U(\chi)$ becomes flat for $\chi\to\infty$.

In a situation with several scalar fields, for example including the Higgs doublet, these statements apply to the relative minimum of the effective potential with respect to the additional scalar fields. In the multifield space the potential $U$ has a flat valley for $\chi\to\infty$. Furthermore, the quantum gravity bound for the maximal increase of the potential for large values of scalar fields suggests that $U$ also becomes flat for asymptotically large values of the Higgs scalar. (This flattening of the effective potential is somewhat analogous to the approach of the effective potential to convexity for spontaneous symmetry breaking~\cite{TETW}.)

The realization of quantum scale symmetry at the UV-fixed point differs from the IR-fixed point. For our ansatz the limit $k\to\infty$ contains the scale $k$ in the effective action~\eqref{1} since $F\sim k^2$, $U\sim k^4$. The action is not invariant under the same transformation of fields as for the IR-fixed point. We may, however use a new \qq{scaling frame}~\cite{CWIQM,CWPFF} with metric
\bel{QS18}
g_{\mu\nu}'=\frac{k^2}{\chi^2}g_{\mu\nu}\ .
\ee
Performing the corresponding Weyl scaling the effective action~\eqref{1} with $F=2w(\chi)k^2$, $U=u(\chi)k^4$ becomes
\begin{align}
\label{QS19}
\Gamma=\int_x&\sqrt{g'}\bigg\{-w\chi^2R'+u\chi^4 \nn\\&
+\frac12\bigg(\frac{\chi^2K}{k^2}-12w-12\frac{\partial w}{\partial\ln\chi}\bigg)\partial^\mu\chi\partial_\mu\chi\bigg\}\ .
\end{align}
This action becomes independent of $k$ if $w$ and $u$ are constant and $K$ is proportional to $k^2/\chi^2$. This is precisely the case for our crossover ansatz~\eqref{CC10} for $\chi\to0$ if $\sigma=2$. The global scale transformation $\chi\to\alpha\chi$, $g_{\mu\nu}'\to\alpha^{-2}g_{\mu\nu}'$ transforms now $g_{\mu\nu}'$, while $g_{\mu\nu}$ remains invariant.

The invariance under a scaling of $\chi$ with fixed $g_{\mu\nu}$ is also visible in the limit $\chi\to0$ of our ansatz in terms of $g_{\mu\nu}$
\bel{QS20}
\Gamma=\int_x\sqrt{g}\bigg\{-w_0k^2R+\frac{8\xi\kappa k^2}{\chi^2}\partial^\mu\chi\partial_\mu\chi+u_0k^4\bigg\}\ .
\ee
This UV-limit is very simple. It is Einstein-gravity with a different value of the Planck mass$\sim k$, a cosmological constant $\sim k^4$ and a free massless scalar field $\tilde{\sigma}=4\sqrt{\xi\kappa}k\ln(\chi/k)$ with a canonical kinetic term. The obvious solution of the field equations for variable gravity in the limit $\chi\to0$ or $\tilde{\sigma}\to -\infty$ is de-Sitter space, with constant Hubble parameter $H^2=u_0k^2/(6w_0)$. This solution is unstable towards increasing small non-zero values of $\chi$. The inflationary epoch is directly linked to this unstable de-Sitter solution.

Quantum scale symmetry at the UV-fixed point gives a strong argument in favor of a divergence of $K$ in the limit $\chi\to0$. The degree of the divergence may be questioned, however. For $K$ increasing with a different power $K\sim(k/\chi)^{\sigma}$, we can define a renormalized scalar $\chi_R$ according to $\ln(\chi_R/k)=(\chi/k)^{1-2\sigma}$. Employing a Weyl scaling which replaces $\chi\to\chi_R$ in eq.~\eqref{QS18} leads to a scale invariant effective action similar to eq.~\eqref{QS19}, with $\chi$ replaced by $\chi_R$. Now the fields $\chi_R\to\alpha\chi_R$, $g_{\mu\nu}'\to\alpha^{-2}g_{\mu\nu}'$ undergo the canonical transformations, translating again to constant $g_{\mu\nu}$. For $\sigma\neq 2$ the transformation of $\chi$ becomes, however, a non-linear transformation. This could suggest that $\sigma=2$ may be singled out for the scaling solution. For a clarification of this issue the gravity induced scalar anomalous dimension should be computed for the limit $\chi\to0$. These arguments concern the asymptotic behavior for $\chi\to 0$, while an effective $\chi$-dependence of $\sigma$ for non-zero $\chi$ remains possible.

\medskip\noindent\textbf{(8) Cosmological scaling solution}
\smallskip

After the end of inflation the Universe enters a \qq{kination} epoch for which the kinetic energy of the scalar field dominates. Realistic cosmology requires entropy production by heating the universe, producing the particles whose energy density dominates in the radiation dominated epoch. During this epoch the energy density of the scalar field may either be negligible, with dynamical dark energy playing a role only later. As an attractive alternative the evolution of the universe enters a cosmic scaling solution~\cite{Wetterich_1988,CLW,LA1} which is a \qq{cosmic attractor} if the parameter $\alpha$ in the kinetial $Z(\vp)$ is sufficiently large. We discuss this cosmic scaling solution in the Einstein frame with a canonical kinetic term for the scalar field $\tilde{\sigma}$ and exponential potential~\eqref{CC12}.

In the limit of constant $\alpha$ the cosmic scaling solution~\cite{Wetterich_1988,CWCMAV} is characterized by a constant fraction of \qq{early dark energy} (EDE)~\cite{EDE1,EDE2},
\bel{CS21}
\Omega_e=\frac n{\alpha^2}\ ,
\ee
where $n=4\ (3)$ for the radiation (matter) dominated epoch. If the IR-fixed point is the conformal fixed point, the function $\alpha(\tilde{\sigma})$ has finally to diverge for $\tilde{\sigma}\to\infty$. Nevertheless, the dependence on $\tilde{\sigma}$ may be smooth enough such that eq.~\eqref{CS21} remains a good approximation. Typically, the recent cosmology is already sufficiently close to the fixed point such that $\alpha$ is large for the recent cosmological epochs. This explains why the EDE-fraction is small. Observation require $\Omega_e$ to be typically below the percent level~\cite{GZAV}. 

For the (approximate) cosmic scaling solution the homogeneous energy density $\rho_h$ of the scalar field decreases at the same rate as the dominant radiation or matter density, $\rho_h=3\Omega_eM^2H^2$. Similarly, the time evolution of the potential obeys
\begin{align}
\label{CS22}
\frac V{M^4}=&\exp\left(-\frac{\alpha\tilde{\sigma}}M\right)=\frac U{F^2}=\frac{uk^4}{\xi^2\chi^4}=\frac{3H^2(1-w_h)\Omega_e}{2M^2}\nn\\
=&\frac{12-2n}n(Mt)^{-2}\ ,
\end{align}
with equation of state parameter $w_h=1/3$ for $n=4$, $w_h=0$ for $n=3$. The large present value of $\chi$ provides for a natural explanation why the present dynamical dark energy density is tiny in Planck units. This is due to the huge age of the universe $M/H\approx10^{60}$, and not to some small intrinsic parameter of the model.

We can employ the frame-invariant quantity~\eqref{CS22} in order to relate different metric frames. For this purpose we also may use the frame invariant Hubble parameter~\cite{CWEU,CWPFVG}
\bel{CS23}
\widehat{\mathcal H}=\mathcal H+\frac{\partial_\eta F}{2F}=a\left(H+\frac{\partial_tF}{2F}\right)\ ,
\ee
with $\eta$ conformal time, $\mathcal H=\partial_\eta\ln a=aH$, and $a$ the cosmic scale factor. In the Einstein frame one has $\partial_tF=0$, while the quantum frame used in eq.~\eqref{1} implies for large $\chi$ and constant $\xi$ the relation $\partial_tF/(2F)=\partial_t\chi/\chi$. For the quantum frame one finds~\cite{CWVG,CWIQM} for the radiation dominated epoch a static universe, $H=0$. In this case the dynamics leading to $\widehat{\mathcal H}\neq0$ is entirely due to the increase of the scalar field. For the matter dominated epoch both the scalar field and the scale factor increase, the latter with a rate different from the Einstein frame, $a\sim t^{1/3}$.

\medskip\noindent\textbf{(9) Cosmon coupling to matter}
\smallskip

Quantum scale symmetry at the IR-fixed point provides for a natural explanation why the coupling of the cosmon to atoms is very weak. No non-linear screening mechanism is necessary in our setting. The strong observational bounds on the time variation of fundamental constants or an apparent violation of the equivalence principle are obeyed naturally. 

Let us denote by $g_{i}$ the dimensionless couplings of the standard model of particle physics. We include in this set the frame-invariant dimensionless mass ratios $h_{0}^{2}/F$ and $\Lambda_{QCD}^{2}/F$, with $h_{0}$ the expectation value of the Higgs doublet (Fermi scale) and  $\Lambda_{Q CD}$ the confinement scale of QCD.
 In the quantum scale invariant standard model~\cite{Wetterich_1988,SZE,GBRSZ,FHNR,SHATK} all $g_{i}$ are independent of the renormalization scale $k$ and therefore of $\chi$. For late cosmology the quantum frame is a scaling frame with $F\sim \chi^{2}$. In this frame one has $h_{0}\sim \chi$ and $\Lambda_{QCD}\sim \chi$, such that all particle masses are proportional to $\chi$. 
The effective Planck mass  $\sim\chi$ increases during the cosmological evolution, and so do all particle masses. Mass ratios and dimensionless gauge couplings or Yukawa couplings remain constant, however, in agreement with the observational bounds. Translating to the Einstein frame with $F=M^{2}$ the constant dimensionless couplings imply now that all particle masses are proportional to $M$.
  For exact quantum scale symmetry there is no coupling of the cosmon $\varphi$ to atoms in the Einstein frame. This implies~\cite{CWQS} the absence of a ``fifth force" due to cosmon exchange, and the absence of a time variation of fundamental couplings despite the time evolution of $\varphi$. This situation is consistent with the role of $\varphi$ as a Goldstone boson that can have at most derivative couplings. 

The IR- fixed point is reached only asymptotically in the infinite future. For a general scaling solution the couplings $g_{i}(\tilde{\rho})$ depend on the dimensionless ratio $\tilde{\rho}=\chi\2 / k\2$. In the Einstein frame this translates to a $\varphi$-dependence of the couplings $g_{i}$ which is, in principle, detectable by an apparent violation of the equivalence principle due to a cosmon mediated fifth force, or by time varying fundamental couplings due to the cosmic evolution of $\varphi$. We will see that for the atoms of baryonic matter this is a very small effect. The overall picture is simple. If at the fixed point for $\tilde{\rho}\to\infty$ the dependence of $g_{i}$ on $\tilde{\rho}$ vanishes, any variation $\tilde{\rho}\partial_{\tilde{\rho}}g_{i}$ will be small for large finite~$\tilde{\rho}$. For present cosmology $\tilde{\rho}\approx 10^{60}$ is huge. 

The general renormalization flow for gauge - or Yukawa couplings depending on $\tilde{\rho}$ and $k$ takes the form
\bel{24}
k\partial_{k}g_{i}=2\tilde{\rho}\partial_{\tilde{\rho}}g_{i}+\beta_{i}(g_{j})\ .
\ee
For $k\2\ll\chi\2$ the metric fluctuations have decoupled and do no longer contribute to $\beta_{i}$. The flow generators $\beta_{i}$ become the standard functional renormalization $\beta$-functions for a model of particle physics without gravity, as obtained by the variation with an infrared cutoff $k$ at fixed $\rho=\tilde{\rho}k^{2}$. For small couplings they coincide with the usual $\beta$-functions, as determined in perturbation theory. Furthermore, the scaling solution is given by a vanishing of the l.h.s. of eq.~\eqref{24}. The couplings depend only on $\tilde{\rho}$ according to 
\bel{25}
\tilde{\rho}\partial_{\tilde{\rho}}g_{i}=-\frac{1}{2}\beta_{i}(g_{j})\ .
\ee

Only the particles with mass $m_{p}^{2}\lesssim k\2$ contribute to the functional renormalization flow. All heavier particles decouple effectively and do no longer contribute to $\beta_{i}$. For $k $ larger than the mass $m_{e}$ of the electron one finds the perturbative running of the fine structure constant. There is therefore a range of $\tilde{\rho}$ for which we expect indeed $\tilde{\rho}$-dependent couplings and the corresponding time variation. This range is given by $\tilde{\rho}\lesssim\tilde{\rho}_{dc}$, with a ``decoupling value" $\rho_{dc}$ determined by
\bel{26}
\frac{m_{e}\2(\chi)}{k\2}=\frac{\chi\2}{k\2	}\bigg{(}\frac{m_{e}}{M}\bigg{)}\2\approx 4\times 10^{-44}\tilde{\rho}_{dc}=1\ .
\ee
The running of the fine structure constant stops, however, for $\tilde{\rho}\gtrsim 10^{45}$ since no more charged particles have mass smaller than $k$. For the cosmic scaling solution~\eqref{CS22} this corresponds to the epoch when the (critical) energy density $\rho_{E}$ in the Einstein frame was larger than $\sim m_{e}^{4}$, 
\bel{27}
\rho_{E}=3M\2H\2>\frac{3u}{\Omega_{e}\xi\2}m_{e}^{4}\ .
\ee
We conclude that for temperatures larger than a ``decoupling temperature" $T_{dc}$, which is roughly in the MeV-range  and depends on $\tilde{u}=u/\xi\2$ and $\Omega_{e}$, the $\tilde{\rho}$-dependence of couplings may indeed lead to a small time-variation of the fine structure constant and other fundamental couplings. This could play a role for the abundance of primordial elements produced during nucleosynthesis~\cite{MSW, VCNS,COC}. 

For the subsequent evolution of the Universe ($T\ll T_{dc}$) the $\tilde{\rho}$-dependence of couplings plays no longer a role. Indeed, for $\tilde{\rho}/\tilde{\rho}_{dc}\gg 1$ the $\tilde{\rho}$-dependence of gauge or Yukawa couplings stops rapidly due to the decoupling of the charged particles. In a minimal setting only the neutrinos, photons and cosmon fluctuations matter in this range. They do not contribute to the corresponding $\beta$-functions. This generalizes to the flow of the ratios $\Lambda_{QCD}(\chi)/\chi$ or $h_{0}(\chi)/\chi$ in the scaling frame. They become independent of $\tilde{\rho}$ for $\tilde{\rho}/\tilde{\rho}_{dc}\gg 1$. In consequence, in the Einstein frame the confinement scale and Fermi scale are independent of $\varphi$. In summary, for the range of $\varphi$ relevant for the present cosmological epoch all renormalizable couplings of the standard model become independent of $\varphi$. Thus $\varphi$ does not couple to atoms. One expects for the present epoch neither a time variation of the renormalizable couplings, nor a fifth force. 

The situation may differ, however, for dark matter if its constituent is a standard model singlet as, for example, a very light scalar field. The renormalization flow in this dark matter sector may induce a non-vanishing dark matter-cosmon coupling for more recent cosmology~\cite{CWCMAV,LACQ}. Neutrino masses arise from non-renormalizable couplings in the standard model. They involve the inverse of mass scales from beyond standard model physics. Without understanding the beyond standard model particle physics a $\varphi$-dependence of neutrino masses or dark matter properties remains an open issue. We will turn to this next.

\medskip\noindent\textbf{(10) Growing neutrino quintessence}
\smallskip

Any realistic dynamical dark energy based on a cosmic scaling solution requires an exit from this solution. Similar to the end of inflation this should be related to some crossover in the coupling functions. The exit from the scaling solution has to occur in a rather recent cosmological epoch for values of $\tilde{\rho}$ close to the present value $\tilde{\rho}_{0}\approx 10^{60}$.
As we have seen before there is no longer any $\tilde{\rho}$-dependence of the renormalizable couplings of the standard model in this range of very large $\tilde{\rho}$. Thus a possible crossover has to be associated to the flow of couplings in the beyond standard model sector. Possible ``portals" are neutrino masses or the dark matter sector. If a cosmic scaling solution plays a role for the radiation and matter dominated epoch the overall picture involves two crossovers in the flow of couplings. The first occurs for small $\tilde{\rho}$ and is associated to the end of inflation. The second occurs for large $\tilde{\rho}$ in the beyond standard model sector, and is associated to the exit from the cosmic scaling solution and the onset of dark energy domination. 

An alternative with possibly only a single crossover are ``thawing quintessence" cosmologies~\cite{Wetterich_1988,HECW,LIN} for which the post-inflationary dynamics drives $\varphi$ to such large values that the potential and kinetic energy density of the cosmon field become negligible during the radiation and matter dominated epochs. The cosmic scaling solution is never reached in this case. Only in the present cosmological epoch the cosmon potential gives a dominant contribution to the energy density of the Universe. This thawing scenario requires for the present range of values of the canonical cosmon field $\tilde{\sigma}$ a very flat potential. Equivalently, $Z(\varphi)$ should be large again for present values of $\varphi$~\cite{HECW,CWCQ}.
We see at present no convincing argument why the scaling solution for $K(\chi)$ should lead to large values of $K$ both for $\chi\to 0$ and $\chi\to \infty$, and small values or even negative values in some intermediate region. If the flow in quantum gravity can exclude such a behavior of $K(\chi)$, the second crossover seems required for a realistic dark energy cosmology. 

A particular interesting candidate is a cosmon coupling to neutrinos as discussed in models of ``growing neutrino quintessence"~\cite{AQCGM,CWGNCS}. Majorana masses of neutrinos involve the inverse of a large mass scale $M_{s}$. The electroweak gauge symmetry allows only neutrino masses $m_{\nu}\sim h_{0}\2/M_{s}$. The small ration $h_{0}/M_{s}$ explains why neutrino masses are much smaller than the masses of charged leptons and quarks which are $\sim h_{0}$~\cite{MIN,YAN,GRS,MACW,LSW}. The scale $M_{s}$ is a characteristic scale of the beyond standard model sector, often related to symmetry breaking of $B$-$L$ (baryon-lepton number) symmetry. 

If $M_{s}$ depends on $\varphi$ in the Einstein frame, the neutrino masses depend on $\varphi$. The resulting cosmon coupling to neutrinos leads to an attractive fifth force between neutrinos in addition to gravity. If $M_{s}(\varphi)$ decreases with increasing $\varphi$ the neutrino masses grow. This can stop effectively the increase of $\varphi$ as soon as neutrinos become non-relativistic, producing on exit from the cosmic scaling solution. Once the increase of $\varphi$ is stopped, the potential $V(\varphi)$ acts very similar to a cosmological constant, with equation of state parameter for quintessence close to -1. Suitable ``growing neutrino quintessence" models based on this simple mechanism seem to be compatible with observation. They lead to an interesting relation between the present dark energy density $\rho_{h}^{(0)}$ and a suitably averaged present neutrino mass $m_{\nu}^{(0)}$\cite{AQCGM}, 
\bel{28}
\Big{(}\rho_{h}^{(0)}\big{)}^{\tfrac{1}{4}}=1.27\bigg{(}\frac{\tilde{\gamma
}(t_{0})m_{\nu}(t_{0})}{eV}\bigg{)}\cdot 10^{-3}eV\ .
\ee
Here $\tilde{\gamma  }$ is a dimensionless quantity characterizing the growth rate of the neutrino masses. The observed present dark energy density is obtained by $\tilde{\gamma}(t_{0})m_{ \nu}(t_{0})=6.15\, eV$, compatible with $\tilde{\gamma}(t_{0})$ of the rough order one for $m_{\nu}(t_{0})$ in the sub-$eV$-range. In other words, the exit from the scaling solution triggered by neutrinos becoming non-relativistic occurs more or less at the right moment for the observed limits on neutrino masses. 

Not much is known at present about the possibility of a suitable crossover in the space of beyond standard model couplings. We will therefore not dwell further on this interesting topic in the present note and refer to refs\cite{AQCGM,CWGNCS,MPRC,ABFPW,CPW}.

\bigskip
In conclusion, we have explored in this note the possible impact of quantum gravity on our understanding of inflation and dark energy. A central point is the scaling solution for the functional flow equations in the presence of metric fluctuations. Its existence is required if gravity can be described by a complete and consistent quantum field theory for the metric and a scalar field. The short distance limit of this scaling solution defines an ultraviolet fixed point which permits the extrapolation of quantum gravity to arbitrarily short distances. The scaling solution has to obey a complex system of non-linear differential flow or renormalization group equations. Its existence and properties place many restrictions on the models used to describe inflation and quintessence. In particular, fundamental scale invariance is a very predictive scheme for a given content of fields or particles. 

The scaling solution fixes the qualitative properties of the effective potential for the scalar field that plays the role of the inflaton in early cosmology and the cosmon for late cosmology. 
Its non-polynomial properties are unfamiliar in perturbation theory. The scaling solution requires an almost constant effective potential $U(\chi)$ for the cosmon, or more generally along the ``cosmon valley" defined by a relative minimum with respect to additional scalar fields. The scaling solution also fixes the qualitative behavior of the field dependence of the coefficient $F(\chi)$ of the curvature scalar. It goes to a constant for $\chi \to 0$, and increases $\sim \chi\2$ for $\chi \to \infty$. 

The still rather limited results on the functional flow in quantum gravity have been found to entail important consequences for quintessential inflation. The most striking feature is a crossover in the effective potential $V(\varphi)$ in the Einstein frame, as shown in Fig.~\ref{fig:1}. This comes in pair with an exponential decrease of $V(\varphi)$ to zero for $\varphi\to\infty$. The latter solves the cosmological constant problem asymptotically for runaway cosmologies for which $\varphi$ increases without bounds towards the infinite future. 

The simple behavior of the scaling solution and the associated quantum scale symmetry solve the issues of ``naturalness'' and ``fine tuning'' for the cosmological constant, the tiny mass of the cosmon and the suppressed cosmon-atom couplings. Small quantities are related to symmetries or dynamics. A possible critical discussion of naturalness should argue why the qualitative behavior of the curves shown in Fig.~\ref{fig:2} is problematic. The Weyl transformation to the Einstein frame obscures the simplicity by introducing  an additional large mass $M$ in the field transformation, which is not a parameter of the quantum field theory.

The high predictivity of fundamental scale invariance puts it in danger to fail. A next important step will be an understanding of the scaling solution for the kinetial $K(\chi)$. If a given model fails to find the increase of $K(\chi)$ to large values for $\chi\to 0$, it would not be compatible with the slow roll behavior during inflation.
(Possible ways to circumvent this statement are a very small value of $u$ that seems not realized by scaling solutions, or Starobinski inflation~\cite{STA} based on a very  large coefficient of the term quadratic in the curvature scale $R\2$.) If it fails to describe the decrease of $K(\chi)$ to small values as $\chi$ increases, it will not allow an end of inflation. And if $K(\chi)$ does not come close to the conformal fixed point for $\chi\to \infty$, there will be too much early dark energy for a cosmic scaling solution to be compatible with observation. 

An approach that can fail at many places is also interesting: it can be tested. Finding the required qualitative properties can give some confidence that the approach goes into the right direction. In this case a quantitative study for given particle physics models may predict the observable properties of inflation and dynamical dark energy. In turn, observations of the primordial fluctuation spectrum may place restrictions on the microscopic models that allow to render quantum gravity complete.  It is well conceivable that the rapid crossover in the kinetial required for a realistic primordial fluctuation spectrum hints towards a crossover in a sector beyond a single cosmon coupled to gravity. This crossover could concern pregeometry~\cite{CWPC}, or be linked to the spontaneous breaking of a grand unified symmetry.

The overall picture of cosmology resulting from the scaling solution of quantum gravity is strikingly simple. The infinite past is ``great emptiness''~\cite{CWGE}, a state with unbroken exact scale symmetry and vanishing expectation values of the metric and scalar fields. The inhomogeneous fluctuations dominate. This state is unstable with respect to a slow increase of the expectation value of the metric $g_{\mu\nu}$ and scalar field $\chi$. Once the expectation values dominate, the Universe becomes homogeneous, as described by inflationary cosmology. In the infinite future quantum scale symmetry becomes again exact due to an infrared fixed point. This symmetry is broken spontaneously by a nonzero value of $\chi$, rendering must particles massive while producing a massless Goldstone boson. The renormalization flow between the UV- and IR-fixed points is characterized by two crossovers, which translate to crossovers in the dependence of coupling functions on $\chi$. The first crossover ends inflation. The second crossover triggers the transition to the present dark energy dominated Universe.

\nocite{*} 
\bibliography{refs}
\end{document}